\documentclass[twocolumn]{aastex631}

\usepackage{graphicx}

\newcommand{\mum}{\ifmmode{\rm \mu m}\else{$\mu$m}\fi}

\shorttitle{Dusty stars in the Sextans A}
\shortauthors{Boyer et al.}

\begin{document}

\title{Discovery of SiC and Iron Dust Around AGB Stars in the very Metal-Poor Sextans A Dwarf Galaxy with JWST: Implications for Dust Production at High Redshift}


\author[0000-0003-4850-9589]{M.~L.\ Boyer}
\affiliation{Space Telescope Science Institute, 3700 San Martin Drive,
Baltimore, MD 21218, USA}

\author[0000-0003-4520-1044]{G.~C.\ Sloan}
\affiliation{Space Telescope Science Institute, 3700 San Martin Drive,
Baltimore, MD 21218, USA}
\affiliation{Department of Physics and Astronomy, University of North Carolina,
Chapel Hill, NC 27599-3255, USA}

\author[0000-0001-6652-1069]{A.\ Nanni}
\affiliation{National Centre for Nuclear Research, ul. Pasteura 7, 02-093 Warsaw, Poland}
\affiliation{INAF - Osservatorio astronomico d'Abruzzo, Via Maggini SNC, 64100, Teramo, Italy}

\author[0000-0003-1356-1096]{E.\ Tarantino}
\affiliation{Space Telescope Science Institute, 3700 San Martin Drive,
Baltimore, MD 21218, USA}

\author[0000-0003-0356-0655]{I.\ McDonald}
\affiliation{Jodrell Bank Centre for Astrophysics, Alan Turing Building, University of Manchester, Manchester M13 9PL, UK; Open University, Walton Hall, Kents Hill, Milton Keynes MK7 6AA, UK}

\author[0000-0002-8937-3844]{S.\ Goldman}
\affiliation{Space Telescope Science Institute, 3700 San Martin Drive,
Baltimore, MD 21218, USA}

\author[0000-0002-5797-2439]{J.A.D.L. \ Blommaert}
\affiliation{Astronomy and Astrophysics Research Group, Department of Physics and Astrophysics, Vrije Universiteit Brussel, Pleinlaan 2, B-1050 Brussels, Belgium}

\author[0000-0003-2442-6981]{F.\ Dell'Agli}
\affiliation{INAF- Osservatorio Astronomico di Roma, Via Frascati 33, 00078, Monteporzio Catone, Roma, Italy}

\author[0000-0003-4132-1209]{M.\ Di Criscienzo}
\affiliation{INAF- Osservatorio Astronomico di Roma, Via Frascati 33, 00078, Monteporzio Catone, Roma, Italy}

\author[0000-0002-1693-2721]{D.~A.\ Garc{\'\i}a-Hern{\'a}ndez}
\affiliation{Instituto de Astrof{\'\i}sica de Canarias, E-38205 La Laguna, Tenerife, Spain}
\affiliation{Departamento de Astrof{\'\i}sica, Universidad de La Laguna, E-38206 La Laguna, Tenerife, Spain}

\author[0000-0003-1319-4089]{Robert D. Gehrz}
\affiliation{Minnesota Institute for Astrophysics, School of Physics and Astronomy, 116 Church Street SE, University of Minnesota, Minneapolis, MN 55455, USA}

\author[0000-0003-2723-6075]{M.A.T.\ Groenewegen}
\affiliation{Koninklijke Sterrenwacht van Belgi\"e, Ringlaan 3, B--1180 Brussels, Belgium}

\author[0000-0001-8392-6754]{A.\ Javadi}
\affiliation{School of Astronomy, Institute for Research in Fundamental Sciences (IPM), Tehran, 19568-36613, Iran}

\author[0000-0003-4870-5547]{O.\ C.\ Jones}
\affil{UK Astronomy Technology Centre, Royal Observatory, Blackford Hill, Edinburgh, EH9 3HJ, UK}

\author[0000-0003-2743-8240]{F.\ Kemper}
\affiliation{Institut de Ciències de l'Espai (ICE, CSIC), Can Magrans, s/n, E-08193 Cerdanyola del Vallès, Barcelona, Spain}
\affiliation{ICREA, Pg. Lluís Companys 23, E-08010 Barcelona, Spain}
\affiliation{Institut d'Estudis Espacials de Catalunya (IEEC), E-08860 Castelldefels, Barcelona, Spain}

\author[0000-0001-9910-9230]{M. Marengo}
\affiliation{Department of Physics, Florida State University, Tallahassee, FL 32303, USA}

\author[0000-0001-5538-2614]{Kristen B.~W.\ McQuinn}
\affiliation{Space Telescope Science Institute, 3700 San Martin Dr., Baltimore, MD 21218, USA}
\affiliation{Department of Physics and Astronomy, Rutgers, The State University of New Jersey, 136 Frelinghuysen Rd, Piscataway, NJ 08854, USA}

\author[0000-0002-0861-7094]{Joana M. Oliveira}
\affiliation{Lennard-Jones Laboratories, School of Chemical \& Physical Sciences, Keele University, ST5 5BG, UK}

\author[0000-0002-9300-7409]{Giada Pastorelli}
\affiliation{Padova Astronomical Observatory, Vicolo dell'Osservatorio 5, Padova, Italy}

\author[0000-0001-6326-7069]{Julia Roman-Duval}
\affiliation{Space Telescope Science Institute, 3700 San Martin Dr., Baltimore, MD 21218, USA}

\author[0000-0002-6858-5063]{R.\ Sahai}
\affiliation{Jet Propulsion Laboratory, California Institute of Technology, Pasadena, CA 91109, USA}

\author[0000-0003-0605-8732]{Evan D.\ Skillman}
\affiliation{Minnesota Institute for Astrophysics, University of Minnesota, Minneapolis, MN 55455, USA}

\author[0000-0002-2996-305X]{S.\ Srinivasan}
\affiliation{Instituto de Radioastronom\'ia y Astrofisica (IRyA), Universidad Nacional Aut\'onoma de M\'exico (UNAM), Antigua Carretera a P\'atzcuaro, 8701, Ex-Hda. San Jos\'e de la Huerta, Morelia, Michoac\'an, 58089, M\'exico}

\author[0000-0002-1272-3017]{J.~Th.\ van Loon}
\affiliation{Lennard-Jones Laboratories, School of Chemical \& Physical Sciences, Keele University, ST5 5BG, UK}

\author[0000-0002-6442-6030]{Daniel R. Weisz}
\affiliation{Department of Astronomy, University of California, Berkeley, Berkeley, CA, 94720, USA}

\author[0000-0002-4678-4432]{Patricia A. Whitelock}
\affiliation{South African Astronomical Observatory, PO Box 9, 7935 Observatory, South Africa; Department of Astronomy, University of Cape Town, 7701 Rondebosch, South Africa}

\begin{abstract}

Low-resolution infrared spectroscopy from JWST confirms the presence of SiC and likely metallic iron dust around asymptotic giant branch (AGB) stars in the Sextans A dwarf galaxy, which has a metallicity $\sim$1\%--7\%~$Z_\odot$. While metal-poor carbon-rich AGB stars are known to produce copious amounts of amorphous carbon dust owing to the dredge up of newly synthesized carbon, this is the first time that Si- and Fe-bearing dust has been detected at this extreme metallicity. Of the six AGB stars observed, one is an intermediate-mass ($\sim$1.2--4~$M_\odot$) carbon star showing SiC dust, and another is an oxygen-rich M-type star with mass $\sim$4--5~$M_\odot$ that is likely undergoing hot bottom burning. The infrared excess of the M-type star is strong, but featureless. We tested multiple dust species, and find that it is best fit with metallic iron dust. Assuming its dust-production rate stays constant over the final 2--3$\times$10$^4$~yr of its evolution, this star will produce $\sim$0.9--3.7 times the iron dust mass predicted by models, with the range depending on the adopted stellar mass. The implications for dust production in high-redshift galaxies are potentially significant, especially regarding the assumed dust species used in dust evolution models and the timescale of AGB dust formation. Stars on the upper end of the AGB mass range can begin producing dust as early as 30--50~Myr after they form, and they may therefore rival dust production by supernovae at high redshift.

\end{abstract}


\section{Introduction}
\label{s.intro} 

Thermally-pulsing stars on the asymptotic giant branch (AGB)\footnote{Throughout, the term ``AGB stars" refers to thermally-pulsing AGB stars. Early AGB stars are not considered here.} produce substantial amounts of dust in the final stages of their evolution and are a major source of dust in the interstellar medium (ISM) \citep[e.g.,][]{Gehrz+1989, Tielens+2005, Matsuura+2009, Boyer+2012, Srinivasan+2016}.  The other primary stellar dust source is supernovae (SNe), though it is not yet known whether SNe are net creators or destroyers of dust, since they can destroy both pre-existing dust in the forward shock and their own newly-synthesized dust in the reverse shock \citep[e.g.,][]{Schneider+2024}. The relative dust contribution from AGB stars and SNe has long been under debate. In our own Solar System, presolar grains are primarily of AGB origin \citep{Hoppe2010, Hoppe+2022}, but the AGB contribution is less clear at low metallicity. Studies that model dust evolution in the ISM conflict, with some showing that the AGB contribution is insignificant compared to the contribution from supernovae and grain growth in the ISM \citep[e.g.,][]{Dwek+2011, Rowlands+2014, Michalowski+2015, Nanni+2020, Sawant+2025}, while others show that AGB stars can dominate the dust input \citep[e.g.,][]{Valiente+2009, Zhukovska+2014, DeLooze+2020}. 
Whether or not AGB stars can produce dust at low metallicity has strong implications for dust production in the early Universe.

Metallicity's influence on AGB dust is difficult to model due to complex processes such as mass loss, dredge up, and pulsation \citep[][and references therein]{Habing2004, Karakas2017, Tosi2022}.  Observationally, carbon-rich AGB stars in nearby galaxies have proven to be prolific dust producers down to at least 0.01~$Z_\odot$, and their dust-production rates have weak, if any, dependence on metallicity  \citep{vanLoon+2008, Sloan+2012, Sloan+2016, Boyer+2015b, Boyer+2017}, presumably because they can produce their own carbon and dredge it up to the surface where it can condense into carbonaceous dust.
However, carbon stars are relatively low-mass stars \citep[$\lesssim$4~M$_\odot$;][]{Ventura+2012, Marigo+2017, Karakas+2018} and do not produce dust until a few hundred Myr after their formation. More massive M-type AGB stars (up to $\sim$8~M$_\odot$) evolve much more quickly, potentially impacting galaxies at high redshift. However, these stars experience nuclear burning at the base of their envelope that destroys newly synthesized carbon before it can be dredged up to the surface in a process called hot bottom burning \citep[HBB;][]{sackmann1992, Boothroyd+1993}, so they remain oxygen-rich and rely on heavier metals for dust production such as Si, Mg, and Fe to form silicate-based dust \citep[see e.g.,][]{GH+2006, GH+2007a, gh2007}. Consequently, massive M-type stars should struggle to produce dust at high redshift due to the limited availability of these elements. Observations of stars in the Magellanic Cloudsappear to confirm this expectation \citep[e.g.,][]{vanLoon+2000, Sloan+2008}.

However, photometric data from the Spitzer Space Telescope and the Hubble Space Telescope (HST) showed that massive oxygen-rich AGB stars may produce dust in galaxies with gas-phase metallicities as low as $\sim$3\% solar metallicity \citep{Boyer+2017}. The existence of these stars suggests that AGB stars may indeed contribute significantly to the dust reservoirs seen in high-redshift galaxies.

This work presents spectroscopic data from the JWST \citep{Gardner+2023, Rigby+2023} from program JWST-GO-1619 (PI: Boyer),\footnote{The data can be found in MAST: \dataset[10.17909/5wd9-m828]{https://archive.stsci.edu/doi/resolve/resolve.html?doi=10.17909/5wd9-m828}} which targeted AGB stars in Sextans~A.  Sextans~A is a member of a galaxy association just outside of the Local Group \citep[$\sim$$1.406 \pm 0.038$~Mpc;][]{Yan+2025}. It is among the most metal-poor star-forming galaxies that harbors a sizable stellar population \citep[$4.4\times10^6~M_\odot$;][]{McConnachie+2012} that is spatially resolvable in the infrared with JWST. The AGB stars in Sextans~A have a metallicity between $\sim$1\% and 7\% solar, with the lower bound traced by ancient red giant branch stars \citep[${\rm [Fe/H]} = -1.85$;][]{Sakai+1996} and the upper bound traced by the gas-phase metallicity in the ISM \citep[$12+{\rm log( O/H)} = 7.54 \pm 0.06$\footnote{Assuming a solar oxygen abundance of $12+\rm{log[O/H]} = 8.69 \pm 0.04$
\citep{Asplund+2021}};][]{Skillman+1989, Kniazev+2005} and spectra of 3 massive A-type stars \citep[${\rm [M/H]} = -1.09$;][]{Kaufer+2004}.  Recent studies suggest that the AGB population may have metallicity near $\sim$3\% $Z_\odot$: \citet{Telford+2021} measured 3\% $Z_\odot$ in a massive O-type star in Sextans~A, and McQuinn et al., (in preparation) find that the metallicity of the bulk of the stellar population is between ${\rm [M/H]} = -1.8$ and $-1.5$ (or 1.6--3.2\% solar), based on a fit to the JWST color-magnitude diagram.

An infrared photometric survey of Sextans~A with Spitzer identified several evolved stars with very red [3.6]--[4.5] colors, indicating dust production \citep{Boyer+2015b, Boyer+2017, Goldman+2019}.  This study targets six of these stars with the Low-Resolution Spectrograph \citep[LRS;][]{Kendrew+2015} on the Mid-InfraRed Instrument \citep[MIRI;][]{Wright+2023}.  The resulting spectra cover 5--14~$\mu$m and probe key molecular and dust features in the stellar envelopes.  This program also obtained images of Sextans~A with MIRI and the Near-InfraRed Camera \citep[NIRCam;][]{Rieke+2023}.

In Section~\ref{s.obs}, we describe the observations and data reduction. Section~\ref{s.analysis} describes the resulting spectra, and Section~\ref{s.disc} discusses our findings and the implications at high redshift. We find that, despite the extreme metallicity of Sextans~A, one carbon stars harbors SiC dust, and a massive M-type star likely harbors metallic iron dust.

\section{Observations} \label{s.obs} 

\begin{deluxetable}{lrllrc} 
\tablecaption{The JWST/LRS sample in Sextans A}
\label{t.sample}
\tablehead{
  \colhead{Target\tablenotemark{a}}    & \colhead{Obs.}    & \colhead{RA} & \colhead{Dec.} &
  \colhead{F$_{10~\mum}$} & \colhead{Stellar}\\
   &  &
  \multicolumn{2}{c}{J2000\tablenotemark{b}} &
  \colhead{($\mu$Jy)} & \colhead{Type\tablenotemark{c}}
}
\startdata
90034 &   5 & 152.748154 & $-$4.682967 & 18 & M\\ 
94328 &   7 & 152.741913 & $-$4.717819 & 52 & C\\ 
92104 &  11 & 152.745163 & $-$4.720694 & 73 & C\\ 
86434 &  10 & 152.753311 & $-$4.698519 & 88 & C\\ 
90428 &   9 & 152.747635 & $-$4.713694 & 238 & C\\ 
94477 &   8 & 152.741699 & $-$4.723953 & 177 & C \\ 
\enddata
\tablenotetext{a}{Target names from \citet{Boyer+2015b, Boyer+2017}. In \citet{Jones+2018}, stars 90034 and 90428 have IDs 210 and 17, respectively. The order in this table is based on the F277W--F444W color, see Figure~\ref{f.cmdlm}.}
\tablenotetext{b}{RA and Dec.\ as reported by JWST.}
\tablenotetext{c}{\ This classification is based on the LRS spectra, see section~\ref{s.analysis}.}
\end{deluxetable}

\begin{deluxetable*}{llrrrrrrrrrrr} 
\tablecaption{NIRCam and MIRI Photometry of the LRS Sample}
\label{tab:mags}
\tablehead{
  \colhead{Target} & \colhead{F090W} & \colhead{F140M} & \colhead{F150W} & \colhead{F200W} & \colhead{F277W} & \colhead{F300M} & \colhead{F335M} & \colhead{F444W} & \colhead{F770W} & \colhead{F1000W} & \colhead{F1130W} & \colhead{F1280W} \\
  & \multicolumn{12}{c}{(mag)\tablenotemark{a}}
}
\startdata
90034 & 19.102 & 17.744 & 17.365 & 16.854 & 16.830 & 16.718 & 16.346 & 16.134 & 15.886 & 15.649 & 15.362 & 15.804 \\
& $\pm$0.001 & $\pm$0.001 & $\pm$0.001 & $\pm$0.001 & $\pm$0.001 & $\pm$0.001 & $\pm$0.001 & $\pm$0.001 & $\pm$0.002 & $\pm$0.004 & $\pm$0.011 & $\pm$0.011 \\
94328 & 21.561 & 19.448 & 19.146 & 17.881 & 16.988 & 16.817 & 16.322 & 15.635 & 14.754 & 14.493 & 14.120 & 14.673 \\
& $\pm$0.004 & $\pm$0.002 & $\pm$0.001 & $\pm$0.001 & $\pm$0.001 & $\pm$0.001 & $\pm$0.001 & $\pm$0.001 & $\pm$0.001 & $\pm$0.002 & $\pm$0.005 & $\pm$0.004 \\
92104 & 23.486 & 20.707 & 20.282 & 18.591 & 17.143 & 16.919 & 16.242 & 15.421 & 14.414 & 14.024 & 13.650 & 14.163 \\
& $\pm$0.009 & $\pm$0.004 & $\pm$0.002 & $\pm$0.001 & $\pm$0.001 & $\pm$0.001 & $\pm$0.001 & $\pm$0.000 & $\pm$0.001 & $\pm$0.001 & $\pm$0.003 & $\pm$0.003 \\
86434 & 23.486 & 20.707 & 20.282 & 18.591 & 17.143 & 16.919 & 16.242 & 15.421 & 14.414 & 14.024 & 13.650 & 14.163 \\
& $\pm$0.009 & $\pm$0.004 & $\pm$0.002 & $\pm$0.001 & $\pm$0.001 & $\pm$0.001 & $\pm$0.001 & $\pm$0.000 & $\pm$0.001 & $\pm$0.001 & $\pm$0.003 & $\pm$0.003 \\
90428 & 29.931 & 24.290 & 23.527 & 20.448 & 18.327 & 18.104 & 16.777 & 15.387 & 13.724 & 12.933 & 12.330 & 12.966 \\
& 1.462 & $\pm$0.032 & $\pm$0.012 & $\pm$0.003 & $\pm$0.001 & $\pm$0.002 & $\pm$0.001 & $\pm$0.001 & $\pm$0.001 & $\pm$0.001 & $\pm$0.001 & $\pm$0.001 \\
94477 & \nodata & 28.491 & 27.529 & 23.196 & 19.805 & 19.408 & 17.870 & 16.102 & 14.209 & 13.255 & 12.730 & 13.406 \\
& \nodata & $\pm$0.737 & $\pm$0.176 & $\pm$0.009 & $\pm$0.002 & $\pm$0.003 & $\pm$0.001 & $\pm$0.000 & $\pm$0.001 & $\pm$0.001 & $\pm$0.001 & $\pm$0.001
\enddata
\tablenotetext{a}{Magnitudes are in the Sirius-Vega magnitude system \citep{Gordon+2022}. Uncertainties are listed below each entry and reflect only the noise characteristics reported by DOLPHOT. These are likely underestimated by a factor of 3--10. Artificial star tests that derive more accurate uncertainties will be described in Tarantino et al. (in preparation).}
\end{deluxetable*}

Program JWST-GO-1619 obtained imaging of most of Sextans~A's star-forming disk and LRS spectroscopy of six previously known dusty AGB stars (Table~\ref{t.sample}). This paper focuses on the LRS data. The LRS provides spectra from 5 to $\sim$14~\mum, with a spectral resolving power ($\lambda$/$\Delta$$\lambda$) of $\sim$40 at 5~\mum\ and $\sim$200 at 13~\mum. For each target, we used the FASTR1 readout, with 50 groups and 6 integrations, resulting in 846~s total exposures. The last column of Table~\ref{t.sample} includes the flux density from the LRS spectrum, averaged between 9.8 and 10.2~\mum.

Based on near-infrared photometry from the Hubble Space Telescope \citep{Boyer+2017}, the star Sextans~A 90034 was identified as oxygen-rich with high confidence due to the measurable impact of water vapor absorption on the photometry around 1.4~$\mu$m.  Star 94328 was also identified as potentially oxygen-rich, but with lower confidence because its colors are near the boundary with carbon stars.  Two stars, 92104 and 86434, were classified as carbon stars based on their colors.  The remaining two stars in the sample, 90428 and 94477, were too obscured by dust in the near-infrared to be detected by Hubble.  They were included in the current target list due to their red [3.6]$-$[4.5] colors and variability in the survey of DUST in Nearby Galaxies with Spitzer \citep[DUSTiNGS;][]{Boyer+2015a, Boyer+2015b}.

\begin{figure} 
\includegraphics[width=3.4in]{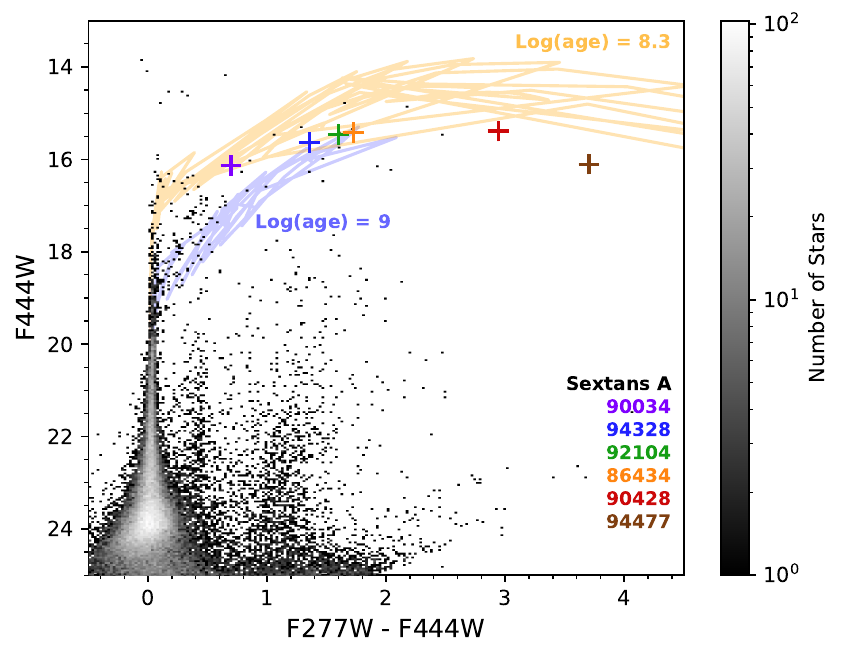}
\caption{A color-magnitude Hess diagram of Sextans A with JWST/NIRCam photometry at 2.7 and 4.4~\mum\ with the positions of the JWST/LRS targets overplotted. {\sc colibri} isochrones with ${\rm [M/H]} = -1.7$  are plotted for log(age) = 8.3 in orange and log(age) = 9 in blue \citep{Marigo+2013}, both showing prominent branches of thermally-pulsing AGB stars at red colors  (F277W$-$F444W $>0$~mag). The population of faint red objects is dominated by unresolved background galaxies and young stellar objects \citep{Warfield+2023, Lenkic+2024}.}
\label{f.cmdlm}
\end{figure}

The spectra were obtained using the standard along-the-slit nod sequence.  The data were processed using the default JWST pipeline \citep{pipeline}\footnote{Using pipeline version 1.15.1 and the CRDS context jwst\_1293.map.}.  We removed NaNs at 6.45 and 6.49~\mum\ and a bad pixel at 12.805~\mum\ from all six spectra.  Five of the six spectra also had a bad pixel at 13.69~\mum, which was also removed.  The exception was Sextans A 94328, which had a bad pixel removed at 13.615~\mum.

Table~\ref{tab:mags} lists preliminary NIRCam and MIRI Vega-based magnitudes measured using the DOLPHOT point-spread function photometry package \citep{Dolphin+2000, Dolphin+2016}, which was adapted for use with JWST by the JWST Stellar Populations Early Release Science project \citep{Weisz+2023, Weisz+2024}. A future paper will present the details of the imaging data and photometry  (Tarantino et al., in preparation). The photometric uncertainties reported here only include the photon-noise characteristics reported by DOLPHOT. Artificial star tests are required to determine the true uncertainties, which tend to be 3--10$\times$ higher in HST data with similar depth and crowding \citep{Williams+2014}. Tarantino et al. (in prep) will include these tests.  Figure~\ref{f.cmdlm} plots the LRS targets on an F277W$-$F444W NIRCam color-magnitude diagram; all six stars are quite red in NIRCam (F277W$-$F444W $>$ 0.7~mag) and follow the theoretical dusty AGB branch. The photometry for the rest of the Sextans~A population shown in Figure~\ref{f.cmdlm} was culled by requiring signal-to-noise $>3$, $-0.26 < {\rm sharpness} < 0.26$, ${\rm crowding} < 0.8$, and a photometry quality flag of 1.

\section{Analysis} \label{s.analysis} 

\begin{figure} 
\includegraphics[width=3.4in]{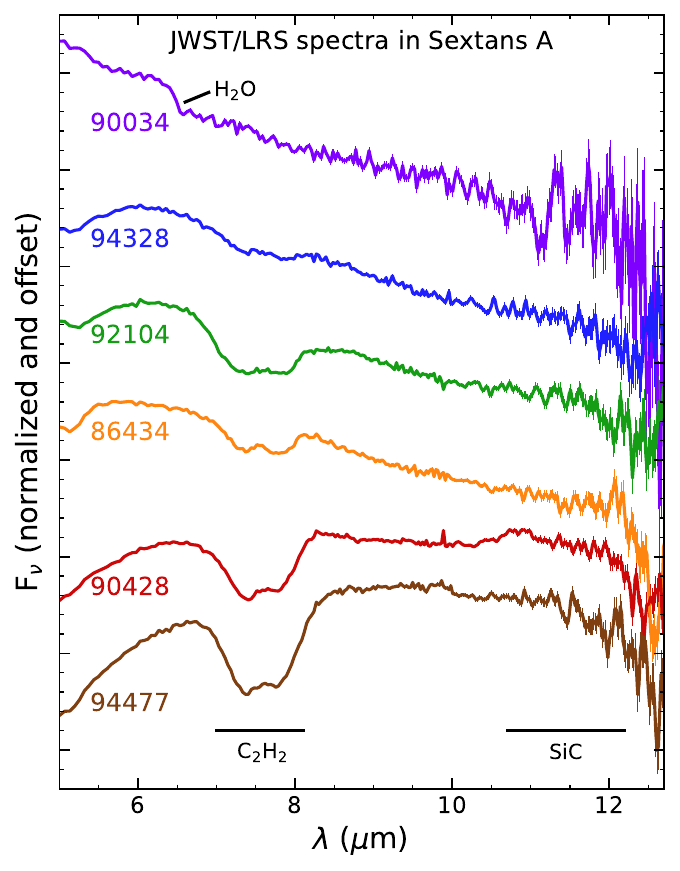}
\caption{The LRS spectra of the six spectroscopic targets in Sextans A.  Uncertainties are plotted, but at the shorter wavelengths are smaller than the width of the plotted spectra. The spectra are smoothed with a 2-pixel boxcar at $\lambda >10~\mu$m. Table~\ref{t.sample} gives the flux densities of the spectra averaged from 9.8 to 10.2~\mum. Spectra are ordered by F277W--F444W color, from top to bottom.}
\label{f.sp6}
\end{figure}

Figure~\ref{f.sp6} presents the LRS spectra of the six targeted stars in Sextans A.  The spectra continue to 14~\mum, but the S/N of the calibrated data are too poor to be of much value beyond the plotted cut-off at 12.7~\mum\ with the current calibration.  

\subsection{Oxygen-rich Spectrum} 

\begin{figure} 
\includegraphics[width=3.4in]{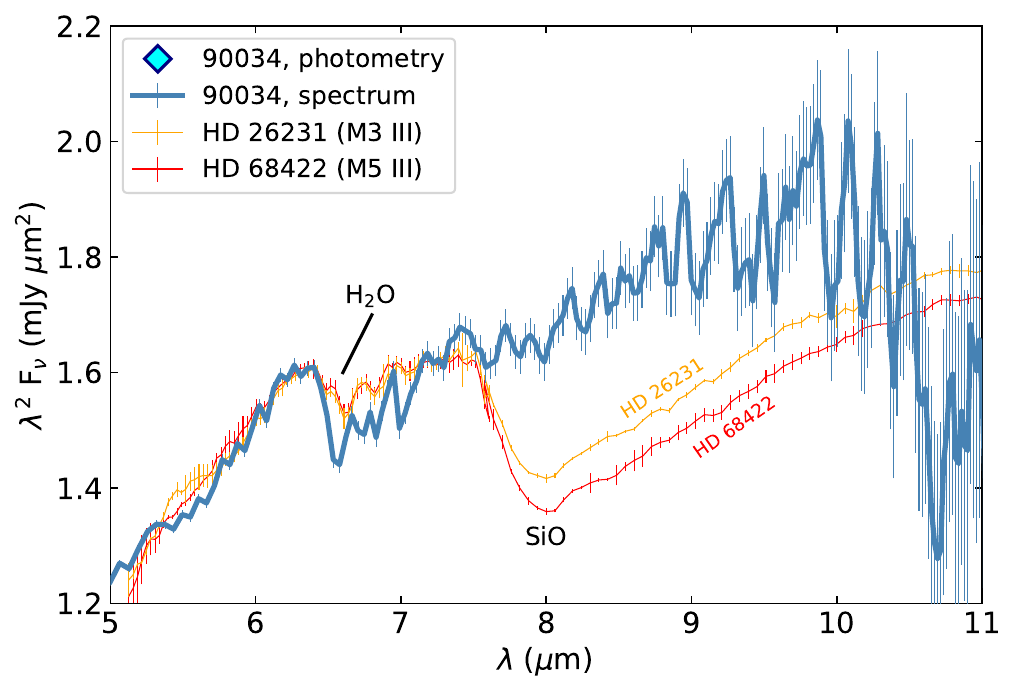}
\caption{The LRS spectrum of Sextans A 90034, plotted in Rayleigh-Jeans units, such that a Rayleigh-Jeans tail would be a horizontal line.  The spectrum is smoothed with a 3-pixel boxcar past 7~\mum\ and a 5-pixel boxcar past 10~\mum.  For comparison, Spitzer spectra of two Galactic M giants \citep{Sloan+2015} are included.  The LRS spectrum shows the same absorption band from water vapor at 6.5~\mum\ apparent in the two M giants, but the silicon monoxide (SiO) band at 8~\mum\ does not appear to be present. The photometric data are from Table~\ref{tab:mags}.}
\label{f.spwater}
\end{figure}

The LRS spectrum of Sextans A 90034 is consistent with its previous HST classification as a high-confidence oxygen-rich candidate \citep{Boyer+2017} and also with an O-rich classification based on a fit to the spectral energy distribution from \citet{Jones+2018}.  It lacks the acetylene (C$_2$H$_2$) absorption at 7.5~\mum\ expected in a carbon star.  The only strong feature in the spectrum in Figure~\ref{f.sp6} is a notch at 6.5~\mum.  Figure~\ref{f.spwater} plots the spectrum of 90034 in Rayleigh-Jeans units to bring out the details largely hidden by the rapid decline to longer wavelengths in $F_{\nu}$ units.
For comparison, Figure~\ref{f.spwater} also plots the spectra of two representative M giants from the Galactic sample of \cite{Sloan+2015}, with clear water and SiO features.  While the spectrum of star 90034 shows water-vapor absorption at 6.5~$\mu$m as expected in a cool oxygen-rich atmosphere, it shows no measurable evidence for SiO absorption around 8~\mum.

We measured equivalent widths following the procedure from \citet{Sloan+2015}, using a Planck function for the continuum tied to 6.2--6.4~\mum\ with temperature 2160$^{+240}_{-190}$K. We find 71.9$\pm$6.8~nm for water vapor and 103.9$\pm$45.2~nm for SiO, or 11.6 and 2.3 sigma, respectively.  We conclude that water vapor is present in the spectrum of 90034, while SiO absorption, if present, is not detectable.

Another possible candidate for the absorption feature at 6.5~\mum\ is silicon monosulfide (SiS), which was identified in Galactic S stars by \cite{Cami+2009, Smolders+2012}.  S stars have a C/O ratio $\sim$1, so the formation of CO has exhausted nearly all the atmospheric C and O, leading to unusual molecular chemistries.  However, we can rule SiS out because its peak absorption would be at 6.7~\mum, not 6.5~\mum. Moreover, water vapor is also supported by the NIRCam photometry, which suggests strong absorption in the F277W and F300M filters that is almost certainly due to water \citep[e.g., see][]{Aringer+2016}.

\subsection{Carbon-rich spectra} 

The remaining five spectra in Figure~\ref{f.sp6} all show an absorption band from acetylene centered at 7.5~\mum, which confirms their carbon-rich nature.  The band is weakest in Sextans A 94328, the star with near-infrared colors from Hubble that placed it close to the boundary between oxygen- and carbon-rich stars.  All five spectra show CO absorption at 5~\mum, and star 90428 has an emission feature at $\sim$11.3~\mum, usually attributed to silicon carbide (SiC) dust \citep{Speck+2009, Sloan+2014}.

To better quantify the strength of the acetylene absorption at 7.5~\mum\ and the SiC dust emission at $\sim$11.3~\mum, we have applied the Manchester Method, which was developed as a standard analysis tool for low-resolution Spitzer spectra of carbon stars. \citep[described in detain by][]{Sloan+2006}.  In outline, it fits line segments over the acetylene absorption band and under the SiC emission feature using standard wavelength stops (applied here with no changes).  The method also determines a color in two narrow bands at 6.4 and 9.3~\mum, which sample the combined ``continuum'' from the star and amorphous carbon in spectral regions relatively free of molecular band absorption or emission from dust features. The [6.4]$-$[9.4] color is redder for objects with more amorphous carbon dust, which is featureless.


\begin{deluxetable*}{llccccr} 
\tablecolumns{7}
\tablewidth{0pt}
\tablecaption{Analysis of carbon-rich molecules and dust in LRS spectra}
\label{t.results}
\tablehead{
  \colhead{Target} & \colhead{[6.4]$-$[9.4]} &
  \multicolumn{2}{c}{C$_2$H$_2$ at 7.5~\mum} &
  \multicolumn{3}{c}{SiC dust / continuum\tablenotemark{a}} \\
  \colhead{Sextans A} & \colhead{(mag)} &
  \colhead{Eq.\ Width (\mum)} & \colhead{$\lambda_c$ (\mum)} &
  \colhead{(ratio)} & \colhead{$\lambda_c$ (\mum)} & \colhead{S/N}
}
\startdata
94328 & 0.280 $\pm$ 0.011 & 0.085 $\pm$ 0.004 &  7.42 $\pm$ 0.04 & (0.162 $\pm$ 0.126) & 11.41 $\pm$ 1.19 & 1.3\\
92104 & 0.395 $\pm$ 0.009 & 0.263 $\pm$ 0.004 &  7.42 $\pm$ 0.01 & ($-$0.011 $\pm$ 0.135) & 12.05 $\pm$ 1.19 & $-$0.8 \\
86434 & 0.405 $\pm$ 0.008 & 0.121 $\pm$ 0.004 &  7.44 $\pm$ 0.03 & (0.219 $\pm$ 0.130) & 11.54 $\pm$ 1.19 & 1.7\\
90428 & 0.822 $\pm$ 0.002 & 0.249 $\pm$ 0.003 &  7.49 $\pm$ 0.01 & 0.139 $\pm$ 0.019 & 11.26 $\pm$ 0.24 & 7.3 \\
94477 & 1.003 $\pm$ 0.006 & 0.323 $\pm$ 0.004 &  7.54 $\pm$ 0.01 & (0.074 $\pm$ 0.073) & 11.28 $\pm$ 1.19 & 1.0
\enddata
\tablenotetext{a}{\ Measurements in parentheses are discounted due to the central wavelength of the apparent feature, as explained in the text.}
\end{deluxetable*}

Table~\ref{t.results} presents the results of the Manchester Method for the five carbon stars.  The reported central wavelengths ($\lambda_c$) define the wavelength that splits the spectral feature in half, and they serve as a check for a given measurement.  All five spectra have prominent acetylene absorption bands with well-defined and reasonable central wavelengths.  One of the five carbon stars has a clearly detected SiC dust emission feature:  Sextans A 90428 (${\rm S/N} = 7.3$).  The others have  $<$2-$\sigma$ SiC detections and highly uncertain central wavelengths, as indicated in the table. 

\begin{figure} 
\includegraphics[width=3.4in]{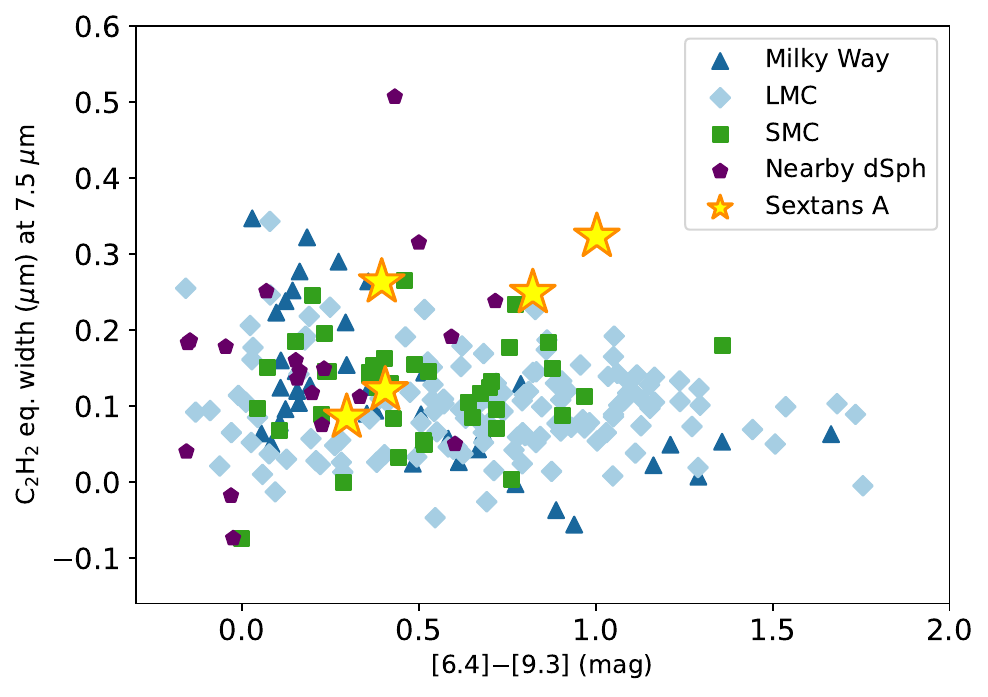}
\caption{The equivalent width of the 7.5~\mum\ acetylene absorption band versus the [6.4]$-$[9.3] color in the LRS and Spitzer/IRS samples \citep{Sloan+2012, Sloan+2016}.}
\label{f.c2h2}
\end{figure}

Figure~\ref{f.c2h2} shows the strength of the 7.5~\mum\ acetylene absorption band as a function of [6.4]$-$[9.3] color, compared to carbon stars observed in the Milky Way, the Magellanic Clouds \citep{Sloan+2016}, and other Local Group dwarf galaxies \citep{Sloan+2012}.  The first take-away from the figure is that the reddest two carbon stars in Sextans A are producing more dust than any of the carbon stars in nearby dwarf spheroidals in the Local Group (purple pentagons), based on their color.  The second is that the strength of the acetylene bands continue the trend seen with metallicity in previous studies.  For those sources with [6.4]$-$[9.3] $>$ 0.5, the acetylene band at 7.5~\mum\ is weakest in the Milky Way, the most metal-rich galaxy examined.  As the metallicity decreases to the Large and Small Magellanic Clouds (LMC and SMC), the strength of the 7.5~\mum\ band generally increases, with the two red carbon stars in Sextans A above all other data points (for [6.4]$-$[9.3] $>$ 0.5). This metallicity trend was also seen in the 3.1~$\mu$m acetylene band in \citet{vanLoon+2008}.

\begin{figure} 
\includegraphics[width=\columnwidth]{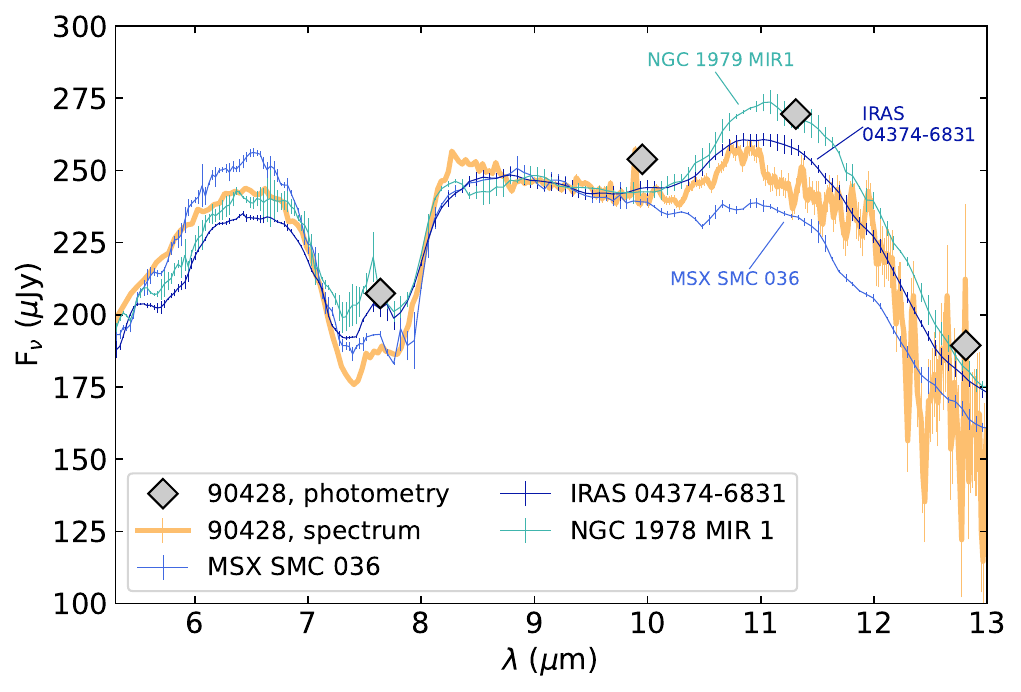}
\caption{The LRS spectrum of Sextans A 90428, compared to the spectra of three similar carbon stars in the Magellanic sample obtained with the Spitzer/IRS, scaled to 9~$\mu$m \citep{Sloan+2016}.  The LRS spectrum is smoothed with a 2-pixel boxcar at $\lambda >10~\mu$m. The photometric data are from Table~\ref{tab:mags}.
}
\label{f.cmp_obs9}
\end{figure}

Figure~\ref{f.cmp_obs9} compares the spectrum of Sextans A 90428, the only candidate in the present sample with SiC emission at 11.3~\mum, to three spectra with similar spectral characteristics ([6.4]$-$[9.3] color, 7.5~\mum\ acetylene strength, and SiC strength) from the Magellanic Spitzer/IRS sample \citep{Sloan+2016}.  While the LRS spectrum is noisier than the comparison spectra, it has the same basic shape, leading us to conclude that this source does show emission from SiC dust in its spectrum.  The photometry, marked with diamonds in Figure~\ref{f.cmp_obs9}, shows a slight offset from the spectrum due to the the limit of the LRS calibration accuracy and the underestimated photometric uncertainties (see \S\ref{s.obs}), but nevertheless also supports the presence of SiC in this target. Notably, among these examples, the Sextans~A acetylene absorption band is comparable or stronger, while the SiC emission feature is weaker. This trend is also apparent in the ISM, where dust depletion decreases with metallicity \citep{Hamanowicz+2024}.

\section{Discussion} \label{s.disc} 

\begin{figure} 
\includegraphics[width=3.4in]{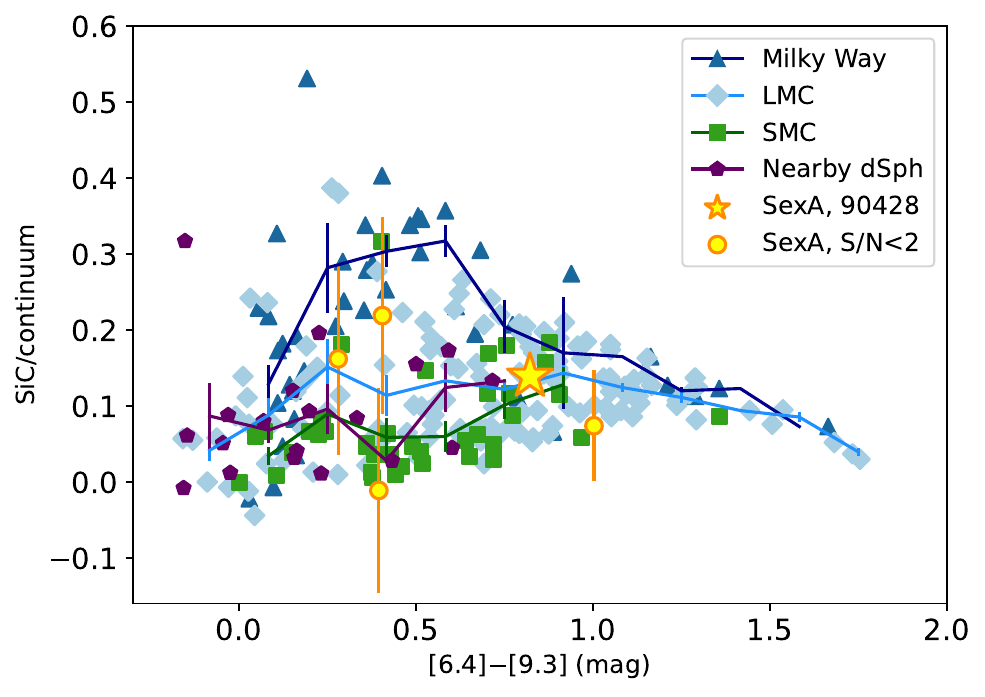}
\caption{The strength of the SiC dust emission feature plotted as a function of the [6.4]$-$[9.3] color from the sample of carbon stars observed in the Magellanic Clouds and other Local Group galaxies with the Spitzer/IRS \citep{Sloan+2012, Sloan+2016}. The Sextans~A source with SiC is plotted as a yellow star and non-detections are plotted as yellow circles. The solid lines show the average SiC strength for each population, as a function of color.}
\label{f.sic}
\end{figure}

\subsection{Carbon Stars and SiC Dust} \label{s.cstars}

Figure~\ref{f.sic} plots the strength of the SiC dust emission as a function of [6.4]$-$[9.3] color, which is a proxy for the dust-production rate in the stars. To put the new Sextans~A sample into context, we also include the same sample of carbon stars from Figure~\ref{f.c2h2}.

Two sequences are apparent in Figure~\ref{f.sic}.  The Galactic carbon stars trace a metal-rich sequence, with rapidly increasing SiC strength as the [6.4]$-$[9.3] color increases to $\sim$0.4, then falling SiC strength from that point on.  \cite{Kraemer+2019} have shown that the bluer sources are associated with weak pulsations in the envelopes of the carbon stars (semi-regular variables), and the redder sources are associated with strong pulsators (Mira variables).  The weak pulsations of the semi-regulars lead to low dust-production rates, and the dust produced is predominantly SiC.  Miras produce much larger quantities of amorphous carbon, which will overwhelm the spectroscopic contribution from the SiC.  The SMC defines a metal-poor sequence, with few stars ever showing strong SiC emission.  The carbon stars in the LMC are split between the metal-rich and metal-poor sequences, and the nearby dwarf spheroidals (Sculptor, Carina, Fornax and Leo I) are almost all on the metal-poor sequence (i.e., weak SiC features). The one carbon star with SiC dust in Figure~\ref{f.sic} is close to the intersection of the two sequences and more aligned with the metal-poor sequence.

\subsection{Dust around the Oxygen-rich Star 90034} \label{sec:mdust}

\begin{figure}
\includegraphics[width=\columnwidth]{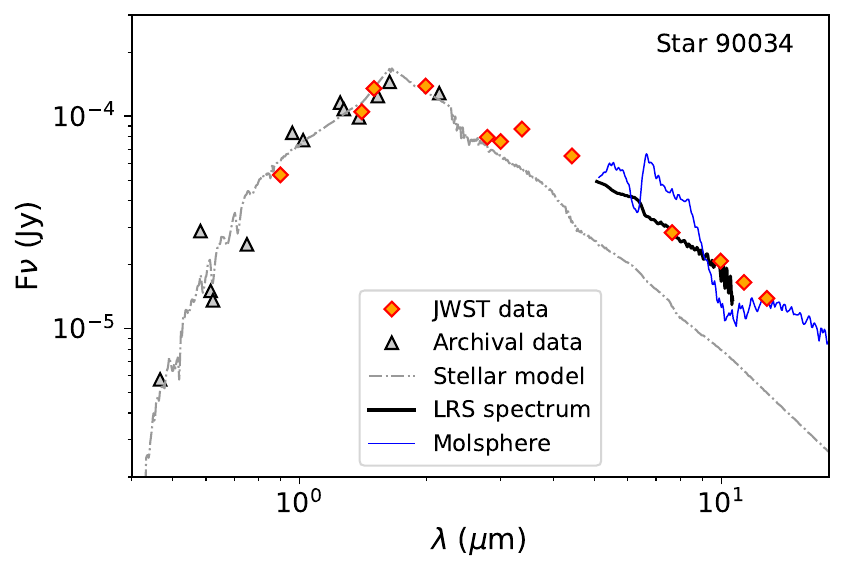}
\caption{The SED of the M-type star 90034.  Gray triangles are archival data, orange diamonds are the JWST data from this program. The gray dash-dot line is the best-fit stellar atmosphere model ($T_{\rm eff} = 3435~K$, ${\rm log}\,L/L_\odot = 4.28$). The SED shows a clear infrared excess over the stellar atmosphere at $\lambda \gtrsim 3$~$\mu$m with no indication of silicate emission at 10~$\mu$m. The absorption around 3~$\mu$m may be due to water vapor in the atmosphere. We also show a water molsphere model (blue line) from \citet{McDonald+2010}, scaled to match the F1280W photometry, see text.}
\label{fig:msed}
\end{figure}

\begin{figure*}
    \includegraphics[width=\textwidth]{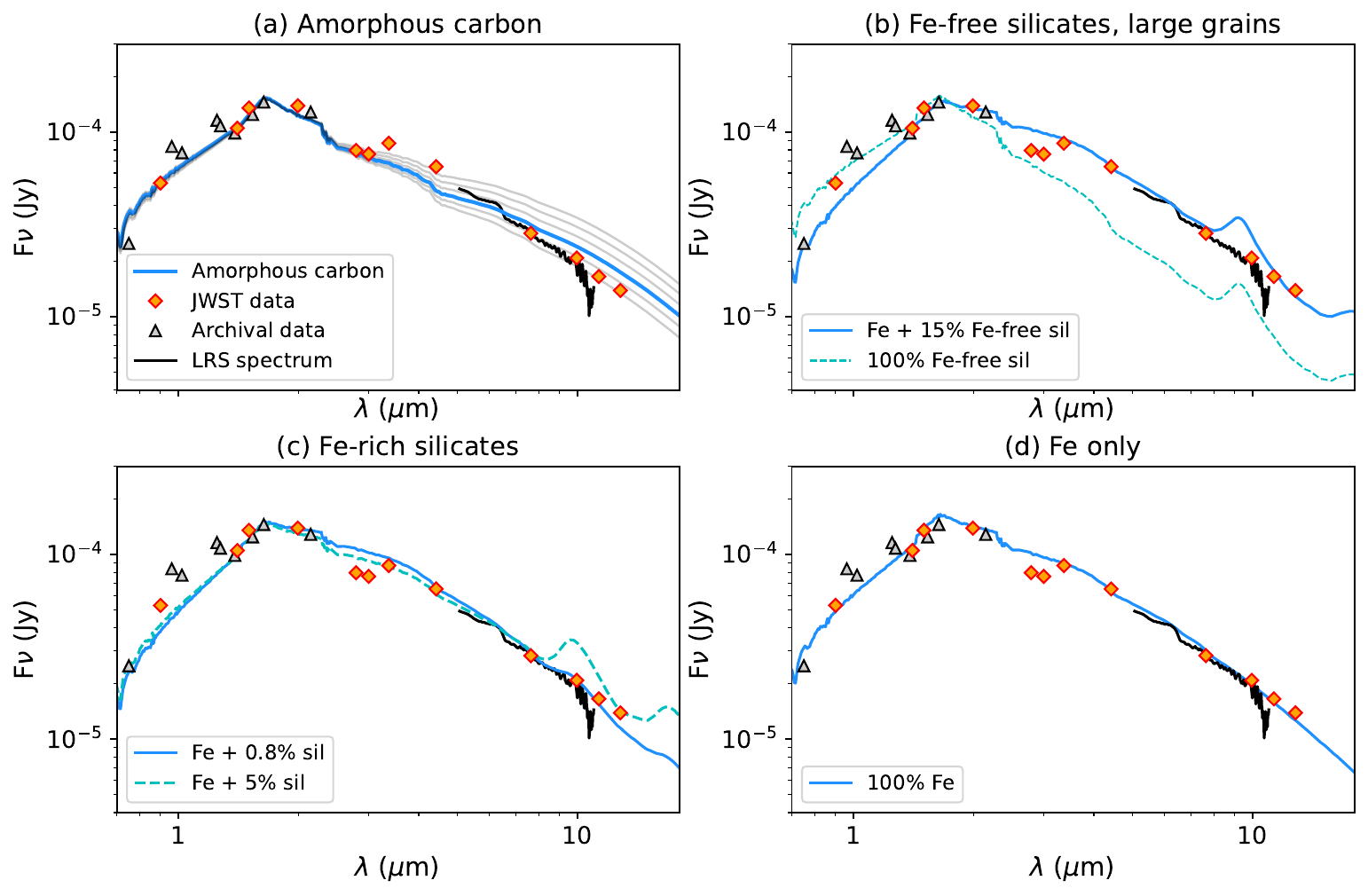}
    \caption{The SED of M-type star 90034, compared to models using different dust species. In panel (a), none of the models fit the data well at $\lambda>3~\mu$m. We show a range of models with differing optical depth in gray. Panel (b) shows a model composed entirely of large silicate grains and a model that includes iron grains to increase the overall infrared excess. Panel (c) shows that only a tiny fraction of silicates ($<$1\%) can be included and still match the overall SED reasonably well. Panel (d) shows a metallic iron model, which matches the data best. See Figure~\ref{fig:zoom} for a zoomed in version of panel (d). }
    \label{fig:SEDs}
\end{figure*}

\begin{deluxetable}{lccl}
\tablecolumns{4}
\tablewidth{0pt}
\tablecaption{Archival Photometry in Figures~\ref{fig:msed} and \ref{fig:SEDs}}
\label{tab:arc}
\tablehead{\colhead{Source} & \colhead{Filter} & \colhead{$\lambda$ ($\mu$m)} & \colhead{F$_\nu$ ($\mu$Jy)}}
\startdata
LBT/LBCB       & g-SLOAN      & 0.47 & \phm{00}5.77$\pm$0.04 \\
PAN-STARRS/PS1 & g$_{\rm P1}$ & 0.48 & \phm{00}0.49$\pm$0.05 \\
GAIA DR3       & G            & 0.58 & \phn28.79$\pm$0.62 \\
LBT/LBCR       & r-SLOAN      & 0.62 & \phn13.58$\pm$0.11\\
PAN-STARRS/PS1 & r$_{\rm P1}$ & 0.62 & \phn15.05$\pm$0.32 \\
PAN-STARRS/PS1 & i$_{\rm P1}$ & 0.75 & \phn24.95$\pm$0.98 \\
PAN-STARRS/PS1 & y$_{\rm P1}$ & 0.96 & \phn83.63$\pm$3.23 \\
Paranal/VISTA  & Y            & 1.02 & \phn77.26$\pm$4.44 \\
Paranal/VISTA  & J            & 1.25 & 116.01$\pm$5.00 \\
HST/WFC3/IR    & F127M        & 1.27 & 108.09$\pm$0.40 \\
HST/WFC3/IR    & F139M        & 1.38 & \phn98.81$\pm$0.36 \\
HST/WFC3/IR    & F153M        & 1.53 & 124.46$\pm$0.34 \\
Paranal/VISTA  & H            & 1.64 & 145.33$\pm$8.55 \\
Paranal/VISTA  & Ks           & 2.14 & 128.53$\pm$8.62 \\
\enddata
\end{deluxetable}

Sextans~A 90034 has red NIRCam colors compared to the red giant branch (Fig.~\ref{f.cmdlm}) indicating dust, but neither the photometry nor the spectrum show evidence of a silicate feature at 10~\mum, and evidence of SiO gas in the atmosphere is marginal (Fig.~\ref{f.spwater}). Instead, the star appears to host a featureless dust continuum in addition to the strong water feature around 6.5\,$\mu$m, A flux deficit at 3\,$\mu$m could also be interpreted as water absorption \citep[cf.][]{Aringer+2016}.

To determine the strength of the infrared excess, we fit a stellar photosphere to the spectral energy distribution (SED). Figure~\ref{fig:msed} shows the SED, incorporating archival data at blue wavelengths from Pan-STARRS1 \citep{panstarrs1}, Gaia Data Release 3 \citep{gaiadr3}, the VISTA Hemisphere Survey \citep{vista}, HST WFC3/IR data \citep{Boyer+2017}, and Large Binocular Telescope (LBT) data from \citet{Bellazzini+2014}, see Table~\ref{tab:arc}. While the JWST data were all taken simultaneously, the archival data were taken at random epochs and are impacted by stellar variability, which can be large in the optical. We have increased the uncertainties to a minimum of 10\% in the archival data to reflect the variability, which corresponds to the low end of the typical optical pulsation amplitude of stars pulsating in the fundamental mode \citep[e.g.,][]{Trabucchi+2017}. To fit the SED, we used a development version\footnote{Version 1.1.dev.20240611, to be described by McDonald et al., in preparation} of PySSED \citep{McDonald+2024} in its default settings, assuming a dereddening of $E(B-V) = 0.034$ mag \citep[][with exinction law from \cite{Fitzpatrick+2019}]{Lallement+2022} and using the default BT-SETTL model atmosphere \citep{Allard+2011}. The fit was restricted to $\lambda<3~\mu$m and returned $T_{\rm eff} = 3440$~K and $L = 19,150~L_\odot$. Stellar evolution models suggest that a star with this luminosity has a mass of $\sim$4--5~$M_\odot$ and is likely undergoing HBB. The dashed gray line in Figure~\ref{fig:msed} shows the best fit photosphere. The JWST data (orange diamonds and black line) show strong, featureless dust continuum emission well in excess of the stellar photosphere at $\lambda \gtrsim 3~\mu$m.

Dust continuum excess has been seen in Galactic stars before (particularly red supergiants), though it tends to appear alongside emission from silicates and has been attributed to either metallic iron \citep[e.g.,][]{Kemper+2002, Marini2019}, amorphous alumina \citep[e.g.,][]{Verhoelst+2006}, amorphous carbon or a water-rich gaseous ``molsphere'' \citep{Verhoelst+2009}. Entirely featureless excesses among oxygen-rich stars have so far been confined to metal-poor, {\em low mass} stars that do not experience HBB \citep[$\lesssim$1~M$_\odot$;][]{Boyer+2009,McDonald+2009,McDonald+2010,McDonald+2011,Sloan+2010}. 

The SED of star 90034 shows strong water bands, especially in the spectrum at 6.5~$\mu$m (Fig.~\ref{f.spwater}), perhaps suggesting the presence of a water-based ``molsphere''. However, \citet{McDonald+2010} showed that a water molsphere could not replicate the overall excess infrared flux in $\omega$\,Cen star V42 without incorporating an additional opacity source such as metallic iron. In Figure~\ref{fig:msed}, we show the water molsphere model from \citet{McDonald+2010}, scaled to the F1280W flux (blue line).\footnote{This molsphere model has column density ${\rm n_{H_2O} = 10^{21} cm^{-2}}$, temperature 1400~K, and is at 3R$_*$ \citep{McDonald+2010}.} It is clear that, in addition to the opacity issue, the molsphere model produces several features that are not present in the photometry or the spectrum of star 90034, and indeed \citet{McDonald+2010} were not able to produce a featureless spectrum using a molesphere. We thus rule out a water molsphere as the primary source of the infrared excess. However, the strong water band visible in the spectrum suggests that water does contribute.



A handful of dust species are capable of producing a featureless dust excess: metallic iron, amorphous carbon, and large silicate grains. We investigate each of these options using stationary wind models coupled with radiative transfer using the {\sc radmc-3d} code \citep{Dullemond+2013}, following the approach from \citet{nanni18,nanni19b}. Details of these models can be found in Appendix~\ref{app:a}. As input, we use the BT-SETTL fitted photosphere model for star 90034, described above. Figure~\ref{fig:SEDs} shows models for each dust species, which we discuss in the following subsections.

\subsubsection{Amorphous Carbon Dust}

Panel (a) of Figure~\ref{fig:SEDs} shows the best-fitting amorphous carbon models. An oxygen-rich star like 90034 should not, in principle, form carbon dust because its atmospheric C/O ratio should be $<$1, leaving no free carbon for dust production. However, a handful of dual-chemistry stars have been identified; for example, RAW\,631 is a carbon star in the SMC that shows evidence of silicate dust, possibly in a circumstellar disk \citep{Jones+2012,Ruffle+2015}. 
According to some stellar evolution models \citep{Marigo+2013, Pastorelli+2019, Pastorelli+2020, Dell-Agli+2019}, C/O in HBB stars may exceed unity towards the end of a star's evolution. In that scenario, it is feasible that carbon dust may be present around an oxygen-rich star like 90034. Amorphous carbon dust could also form through non-equilibrium photo-chemistry, though, in this case, carbon-bearing molecular bands (e.g., C$_2$H$_2$) would be expected in the spectrum of the star.  Despite the possibility of forming amorphous carbon dust, it is clear in Figure~\ref{fig:SEDs}a that the amorphous carbon models do not provide a good fit to the observed SED and spectrum of star 90034. The models approximately match the overall level of infrared excess, but they significantly underestimate the flux at $\lesssim$5~$\mu$m and/or overestimate the flux at $\gtrsim$6.5$\mu$m.  We therefore rule out amorphous carbon dust as a dominant dust species around star 90034.

\subsubsection{Silicate Dust}

Large silicate grains can also suppress the 10~$\mu$m emission feature. According to the models by \citet{Hofner2008}, Fe-free silicate grains (e.g., forsterite, Mg$_2$SiO$_4$) can drive a wind only if they are somewhat large ($a_{\rm gr}\sim0.3~\mu$m).  Panel (b) of Figure~\ref{fig:SEDs} shows our best-fit model using large Fe-free silicate grains ($a_{\rm gr}$ = 0.4~$\mu$m) combined with metallic iron. Metallic iron is needed to increase the overall IR excess $>$3~$\mu$m, as is shown by a model with 100\% silicate grains (dashed cyan line). Even with just 13\% of the dust attributed to Fe-free silicates (solid blue line), the 10~$\mu$m feature is prominent and deviates significantly from the observed spectrum. Very large silicate ($\sim$50~$\mu$m) grains could further suppress the 10~$\mu$m emission feature, but it is difficult to grow very large grains in metal-poor environments due to the limited availability of heavy elements, and indeed our models are not able to grow such large grains for this star. We thus rule out large Fe-free silicate grains.

Panel (c) of Figure~\ref{fig:SEDs} shows a model that includes Fe-rich silicates (olivine and pyroxene), with smaller grain sizes near $a_{\rm gr}\sim0.08~\mu$m. With the addition of just 5\% of this silicate dust, the 10~$\mu$m feature is pronounced. To reasonably match the data, these models must include $<$1\% silicates. 
We conclude that if any silicate dust is present, its contribution to the dust mass is negligible.

It is clear that, even if a small amount of silicate dust is present around star 90034, the overall infrared excess at $\lambda>3~\mu$m cannot be reproduced without the addition of metallic iron dust.

\begin{figure}
    \includegraphics[width=\columnwidth]{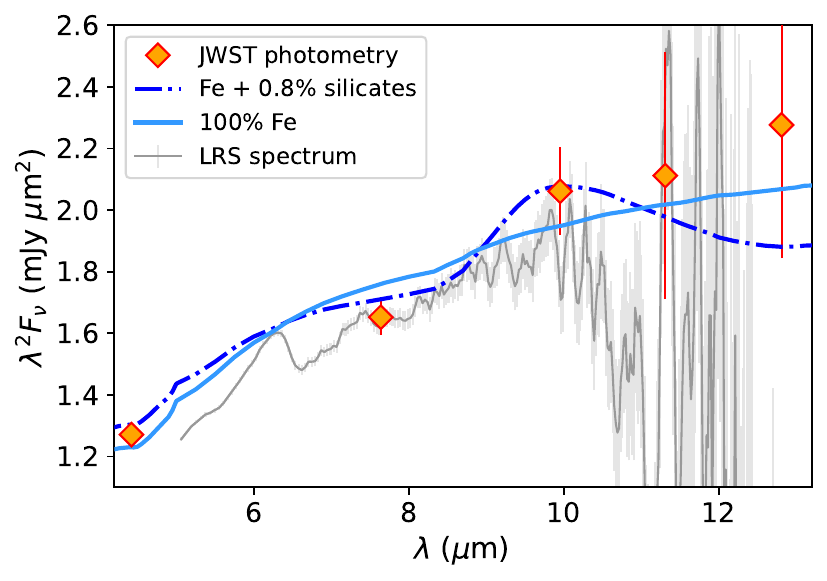}
    \caption{The SED of star 90034, zoomed in to show the details of the spectrum, in Rayleigh-Jeans units. Here we plot 3-$\sigma$ error bars on the photometry to better reflect the true photometric uncertainties (see \S\ref{s.obs}). Both models shown here match the data well, with the 100\% metallic iron model showing a better match to the 12.8~$\mu$m photometry.}
    \label{fig:zoom}
\end{figure}

\subsubsection{Metallic Iron Dust}

Panel (d) in Figures~\ref{fig:SEDs} shows the best-fit model using 100\% metallic iron dust.  The modeled SED shape closely mimics the data and provides the observed featureless spectrum.  Figure~\ref{fig:zoom} zooms in on the spectrum and shows that, while the metallic iron model matches the photometry at 12.8~$\mu$m, the model that includes 0.8\% silicate dust from Figure~\ref{fig:SEDs}c cannot be ruled out, given the noise in both the photometry and spectrum at $\lambda>10~\mu$m.  Based on the models in Figures~\ref{fig:SEDs} and \ref{fig:zoom}, we conclude that metallic iron dominates the dust around star 90034.

The best-fit metallic iron model has an optical depth of $\tau_{\rm V} = 0.6$ in the V-band, a total mass-loss rate of $\dot{M} = 1\times10^{-4}~M_\odot {\rm yr}^{-1}$, and a dust-production rate (DPR) of $\dot{D}_{\rm Fe} = 8\times10^{-10}~M_\odot {\rm yr}^{-1}$. This DPR is on the upper end of the distribution of DPRs in LMC for M-type AGB stars \citep[see Fig.~16 from ][]{Riebel+2012}. This star thus appears remarkably dusty, given its low metallicity, and a DPR this large suggests that this star is near the end of its evolution on the AGB. We note, however, that adopting different grain properties \citep[such as elongated iron grains;][]{McDonald+2011c} and/or different dust optical constants could decrease the DPR.

Since the star is a relatively massive AGB star, we adopted a metallicity of about ${\rm [Fe/H]} = -1.6$~dex for our model, near the high end of the observed CMD distribution (McQuinn et al., in preparation) and the value measured for a massive O-type star in Sextans~A \citep{Telford+2021}. If the wind is pure metallic iron, then the star's iron abundance limits the wind to a gas-to-dust mass ratio of $\psi \gtrsim 22\,000$, assuming the solar iron number density is log(Fe)$_\odot$ = 7.52. Here, the gas and dust mass-loss rates indicate $\psi$ is 5.6$\times$ higher than the limit, indicating that a significant fraction of iron remains in the gas phase. This value is similar to the dust-to-gas ratios derived from depletion measurements in Sextans~A at low dust column densities, where ISM dust growth is minimal \citep{Hamanowicz+2024}.

Our model assumes a constant wind speed of $v_{\rm 0} = 2$~km\,s$^{-1}$, which is consistent with the velocities seen in stars with similar SEDs. In the specific cases of 47 Tuc V3 \citep{McDonald+2019} and the Galactic halo star RU Vul \citep{McDonald+2020}, presence of a featureless IR excess has been linked by sub-mm observations to a wind with a very slow expansion velocity ($v_{\rm exp}$) of only a few km\,s$^{-1}$. Velocities this low suggest that dust driving is ineffectual and that the mass loss may instead be primarily dependent on pulsations \citep[e.g.,][]{McDonald+2016, McDonald+2018}. If, however, a small amount of silicate dust is present in star 90034 (Fig.~\ref{fig:zoom}), it could help to drive a wind if the grains are large enough \citep[$a_{\rm gr} \sim 0.3~\mu$m;][]{Hofner2008}.

\subsection{Dust Production at Low Metallicity} \label{sec:dust}

Figure~\ref{f.cmdlm} shows that all six stars are quite red in F277W$-$F444W, suggesting that dust formation is efficient even at near primoridal metallicity. For the carbon stars, this is not surprising. Even at the low metallicity of Sextans\,A, carbon stars can dredge up newly synthesized carbon to the surface, where it can form amorphous carbon dust.  What is more surprising is the confirmed detection of SiC. Previously, the most metal-poor star with an SiC detection was in Leo\,I \citep{Sloan+2012}, which has a metallicity ranging from [Fe/H] $\approx$ $-1.42\pm0.36$, based on measurements of iron lines \citep{Kirby+2011}, to [Fe/H] = $-1.2\pm0.2$ from calcium triplet measurements in red giant branch stars. Sextans\,A is therefore the most metal-poor galaxy known to harbor stars that can produce SiC dust.

Models predict that SiC dust formation depends strongly on metallicity \citep[e.g.,][]{Zhukovska+2008, nanni14, nanni16, Ventura+2012, Ventura+2018, Dell-Agli+2019, Schneider+2024}, but Figure~\ref{f.sic} shows that star 90428 and a few other stars in nearby metal-poor dSph galaxies \citep{Sloan+2012} have SiC strengths similar to stars in more metal-rich galaxies like the LMC ($\sim$50\%~$Z_\odot$).  Since the sample size is small, it is unclear if these stars are typical or if they are outliers.

For star 90034, metallic iron dust appears to dominate. Metallic iron dust has been invoked to reproduce the spectra of silicate-producing AGB stars where an additional opacity source is required to increase the overall continuum dust excess \citep[e.g.,][]{Kemper+2002, Jones+2014, Marini2019}. In the case of \citet{Marini2019}, the spectra of seven LMC stars show that iron dust accounts for 70--80\% of the total dust mass, counter to the common assumption that massive oxygen-rich stars primarily form silicate dust. For stars more massive than 3--4~M$_\odot$, HBB decreases the surface abundances of magnesium and oxygen, the latter causing a decrease in H$_2$O and SiO molecules. These molecules, together with magnesium, are required to form olivine (Mg$_2$SiO$_4$). \citet{Marini2019} propose that if the metallicity is low enough ($Z \lesssim 10^{-3}$), the star will experience a period where silicates cannot form and iron dust dominates.  The lack of SiO in the spectrum of star 90034 supports this scenario.

Since we are observing only a snapshot of star 90034 during its evolution, it is difficult to estimate how much iron dust it could produce over its lifetime.  The {\sc colibri} stellar evolution models \citep{Marigo+2013, Pastorelli+2019, Pastorelli+2020} predict that stars with mass 4--5~M$_\odot$ reach a mass-loss rate of $\sim10^{-4.5}$--$10^{-3.5}~{\rm M_\odot yr^{-1}}$ in the final 2--3$\times10^4$~yr of their evolution. If we assume a constant mass-loss rate over that time frame and assume that the dominant dust species is also constant, then star 90034 would produce 0.8--$2.4\times10^{-5}$~M$_\odot$ of metallic iron dust. That value is a factor of 0.9$\times$--3.7$\times$ the predicted iron dust mass for a 4--5~M$_\odot$ star with $Z = 3\times10^{-4}$ from \citet{Dell-Agli+2019}.  While the lower end of this range provides good agreement between the models and the data, the higher end would bring the dust mass in line with predictions for more metal-rich models ($Z = 1\times10^{-3}$) and would have a strong impact on galaxy dust budgets. At the same time, substantial iron dust production would agree with depletion-derived ISM dust compositions seen in the SMC, where carbon and iron dust dominate in regions with low column density \citep[i.e., where there is minimal grain growth;][]{Roman-Duval+2022}.  A larger population of metal-poor massive stars is needed to draw definitive conclusions, which may be found in other nearby metal-poor dwarf galaxies like Sextans~B or Sag~DIG \citep{Saviane+2002, Lee+2006}, or more distant metal-poor galaxies like I Zw 18 or DDO 68 \citep{Aloisi+2007,sacchi+2016, Alec+2024}.

The efficient formation of metallic iron dust at this extreme metallicity would potentially have major implications for dust formation at high redshift. Most AGB dust models predict that the amount of metallic iron dust formed by AGB stars at any metallicity is insignificant compared to amorphous carbon and silicate dust \citep{Zhukovska+2008, nanni14, nanni16, Ventura+2012b, Ventura+2018}, so metallic iron is typically not included in cosmic dust evolution models.  While carbon dust production is mostly independent of metallicity \citep[e.g.,][]{Ventura+2012}, silicate dust requires pre-existing metals (Si, Mg) and is expected to form in only very small quantities at low metallicity. Since carbon stars are low mass ($<$3--4~M$_\odot$) and therefore slow to evolve, dust evolution models predict that metal-poor AGB populations do not begin to match supernova dust injection rates until $z \sim 5$ \citep[e.g.,][]{Schneider+2024}. This delayed dust injection precludes a contribution from AGB stars to the massive dust reservoirs that are now being detected in galaxies at redshifts as high as $z \sim 8$ \citep{Witstok+2023b,DEugenio+2024}. If, on the other hand, metal-poor massive AGB stars up to 8~M$_\odot$ are able to efficiently produce metallic iron dust as seen in star 90034, AGB stars could contribute dust as early as 30--50~Myr after they form, or $z > 15$ \citep[e.g., see][]{Schneider+2024}. This would affect not only the dust mass in the ISM at high redshift, but also the dust composition and opacity.

\section{Conclusions}

We present MIRI/LRS spectra of six TP-AGB stars in the metal-poor dwarf galaxy Sextans\,A: five carbon stars, and one M-type star. We find:

\begin{itemize}
    \item One of the five carbon stars (90428) shows evidence of SiC dust at $\sim$11.3~$\mu$m. The strength of the SiC feature is similar to that seen in the more metal-rich SMC and LMC galaxies, while C$_2$H$_2$ absorption is stronger.
    \item The M type star (90034) shows deep water absorption at 6.5~$\mu$m and a strong featureless infrared excess with no obvious silicate emission at 10~$\mu$m, similar to low-mass M-type stars in globular clusters. However, this star's luminosity indicates an initial stellar mass of 4--5~$M_\odot$, placing it on the upper end of the AGB mass distribution where stars are undergoing HBB.
    \item The observed SED of M type star 90034 is best reproduced with 100\% metallic iron dust, with a dust-production rate as high as those seen in the LMC. However, small amounts of silicate dust ($<$1\%) cannot be ruled out. Other potential dust species, including amorphous carbon dust and large silicate grains, resulted in poor fits to the SED.
\end{itemize}

At just $\sim$1--7\% solar metallicity, these stars are among the most metal-poor with confirmed dust production. The presence of SiC in one carbon star confirms that silicon is available for dust production even at these extreme metallicities. The M-type star appears to be producing primarily metallic iron dust with a remarkably high dust-production rate, suggesting that metallic iron dust may be common at low metallicity.

AGB stars with masses 4--5~$M_\odot$ progress to the dust-producing phase on a rapid timescale ($\sim$100~Myr). Assuming star 90034 is representative of other metal-poor HBB stars with masses up to $\sim$8~$M_\odot$, the timescale of dust injection by AGB stars is as early as $\sim$30~Myr after the onset of star formation. Currently, the two most distant galaxies have spectroscopically-confirmed redshifts $>$14. While galaxy MoM-z14 appears to have formed most of its stellar mass in the last 10~Myr \citep{Naidu+2025}, the star-formation history of galaxy JADES-GS-z14-0 suggests its stellar population has been evolving for $\sim$100~Myr \citep{Carniani+2024}. The AGB population in JADES-GS-z14-0 may therefore already be producing Fe-rich dust.

\begin{acknowledgements}

We thank the referee for their helpful input to this manuscript.

This work is based on observations made with the NASA/ESA/CSA James Webb Space Telescope. The data were obtained from the Mikulski Archive for Space Telescopes at the Space Telescope Science Institute, which is operated by the Association of Universities for Research in Astronomy, Inc., under NASA contract NAS 5-03127 for JWST. These observations are associated with program JWST-GO-1619.


We also include observations obtained as part of the VISTA Hemisphere Survey, ESO Program, 179.A-2010 (PI: McMahon) and data from the Pan-STARRS1 Surveys (PS1) survey \citep{Chambers+2016, panstarrs1}

R.S.'s contribution to the  research described here was carried out at the Jet Propulsion Laboratory, California Institute of Technology, under a contract with NASA, and funded in part by NASA via a JWST GO award.
D.A.G.H. acknowledges support from the State Research Agency (AEI) of the Spanish Ministry of Science, Innovation, and Universities (MICIU) of the Government of Spain, and the European Regional Development fund (ERDF), under grant PID2023-147325NB-I00/AEI/10.13039/501100011033. This publication is based upon work from COST Action CA21126 - Carbon molecular nanostructures in space (NanoSpace), supported by COST (European Cooperation in Science and Technology).
A.N.\ acknowledges support from the Narodowe
Centrum Nauki (NCN), Poland, through the SONATA BIS grant UMO-2020/38/E/ST9/00077.
F.K.\ acknowledges support from the Spanish Ministry of Science, Innovation and Universities, under grant number PID2023-149918NB-I00. This work was also partly supported by the Spanish program Unidad de Excelencia María de Maeztu CEX2020-001058-M, financed by MCIN/AEI/10.13039/501100011033.
O.C.J.\ has received funding from an STFC Webb fellowship.
RDG was supported, in part, by the United States Air Force.
P.A.W.\ acknowledges support from the South African National Research Foundation (NRF).
\end{acknowledgements}

\facilities{JWST(NIRCam, MIRI)}

\software{Astropy \citep{astropy:2013, astropy:2018, astropy:2022},
          {\sc dolphot} \citep{Dolphin+2000, Dolphin+2016}, JWST Pipeline \citep{pipeline}, RadMC-3D \citet{Dullemond+2013}, PySSED \citep{McDonald+2024}
          }

\appendix
\section{SED Dust Models}
\label{app:a}

To assess the most likely dust species present around star 90034, we used SED models obtained by combining stationary wind models and radiative transfer, described in detail by \citet{nanni13, nanni14}. 
The wind models use a revised version of the dust formation description from \citet{FG+2006}. A grid of input stellar parameters (effective temperature, luminosity, stellar mass, mass-loss rate) were chosen based on the TP-AGB stellar evolution tracks computed using the {\sc parsec} code coupled with the {\sc colibri} code \citep{Bressan+2012, Marigo+2013}. To compute the dust growth of each species, we assumed an atmospheric C/O ratio (${\rm C}/{\rm O}=1.1$ for carbon dust and ${\rm C}/{\rm O}=0.5$ for all other dust species) and an initial metallicity of Z=0.0004. The input spectrum for each calculation is the best-fit BT-SETTL photosphere model described in section~\ref{sec:mdust}.

We followed the growth of corundum (Al$_2$O$_3$), olivine (Mg$_{2x_{\rm ol}}$Fe$_{2(1-x_{\rm ol})}$SiO$_4$), pyroxene (Mg$_{x_{\rm py}}$Fe$_{(1-x_{\rm py})}$SiO$_3$), quartz (SiO$_2$), periclase (MgO), and metallic iron for the oxygen-rich case, and amorphous carbon for the carbon-rich case. Here, $x_{\rm ol, py}={\rm Mg}_{\rm ol, py}/({\rm Mg}_{\rm ol, py}+{\rm Fe}_{\rm ol, py})$ is the fraction of magnesium in the grain over total magnesium and iron.

For olivine and pyroxene, we assumed that the destruction mechanism is free evaporation. In this case, the destruction rate is computed in analogy to Eq.~3 of \citet{kobayashi11} and outlined in \citet{nanni13, nanni14}. For corundum, quartz and periclase, we assumed the reaction between H$_2$ molecules and the grain surface, (chemisputtering), to be fully efficient. The sticking coefficients $\alpha_i$ (the probability for an atom or molecule to stick on the grain surface), are selected following \citet{Gail99} and \citet{nanni13}. The exceptions are olivine and pyroxene, for which we chose a sticking coefficient of 0.4 to match the observed expansion velocities for a sample of Galactic Miras \citep{Uttenthaler24}.

We assumed a constant wind speed, $v_0=2$\,km\,s$^{-1}$, and selected the seed particle abundance ($\epsilon_s$) to match observations of carbon stars \citep{nanni16, nanni19a} and oxygen-rich Miras \citep{Uttenthaler24}.  The seed particle abundance affects how much dust can form and the grain size. For the iron-rich silicate case, we considered $\epsilon_{\rm s, sil}=10^{-15} \frac{Z}{Z_\odot}$, with both $\epsilon_{\rm s, Fe}=10^{-14} \frac{Z}{Z_\odot}$ and $10^{-13}\frac{Z}{Z_\odot}$ for the included metallic iron. For iron-free silicates, we used $\epsilon_{\rm s, sil}=5\times10^{-17}\frac{Z}{Z_\odot}$, with $\epsilon_{\rm s, Fe}=10^{-13} \frac{Z}{Z_\odot}$. For the 100\% metallic iron case, we use $\epsilon_s=10^{-14}\frac{Z}{Z_\odot}$. In all cases, we assume $Z_\odot = 0.01524$. For amorphous carbon dust, we use $\epsilon_{\rm s, C}=10^{-12}\ (\epsilon_C - \epsilon_O)$, where $\epsilon_C$ and $\epsilon_O$ are respectively the abundances of carbon and oxygen atoms relative to hydrogen. In the oxygen-rich cases, we find that the amount of quartz, periclase, and corundum produced is negligible.

The output of the growth code is used as input for the  \textsc{radmc-3d} radiative transfer code. The approach adopted is analogous to \citet{nanni18, nanni19b}. As input, we used the same BT-SETTL dust-free photospheric spectrum used for the dust-growth calculation and the density profile for each dust species.  Dust temperature profiles for each dust species are directly computed by the {\sc radmc-3d} code.  Following \citet{nanni18, nanni19a}, we adopted the optical constants from \citet{dorsch95} with $x_{\rm ol}=x_{\rm py}=0.05$, \citet{Jaeger94} for iron-free silicates, \citet{ordal88} for metallic iron, and \citet{Jaeger98} for carbon dust (sample at 1000 C).
The dust opacities are consistently computed for the typical grain size obtained at the end of each integration.

\bibliography{main}{}
\bibliographystyle{aasjournal}

\end{document}